# Towards a Grid-based approach to Traffic Routing in VANET

Noman Islam, Zubair A. Shaikh, Shahnawaz Talpur
Center of Research in Ubiquitous Computing
National University of Computer & Emerging Sciences
Karachi, Pakistan

*Abstract*—**Vehicular Ad hoc Network (VANET) is a wireless network that is formed on the fly between a collection of cars connected by wireless links. VANET has gained a great amount of attention during recent years and is used for a large number of safety (accident warning, weather notification) and non-safety (multimedia, gaming) applications. One of the most promising areas where VANET can be proved to be very useful is Traffic Management. Various approaches that have been tried for Traffic Management ranges from those that takes assistance from roadside infrastructure [7], [8] to purely infrastructure-less approaches [6]. In this paper we have addressed a very important issue of Traffic Management. Our proposed approach provides assistance in traffic routing by integrating VANET and Grid Computing. By using our approach a driver can always find out a feasible route to reach its destination based on congestion all around the path from its current position to destination. The proposed approach is scalable, simple and can be easily adapted to run in any environment.**

*Keywords- VANET, Traffic Management, Grid Computing, Traffic Routing, Ad hoc Network,*

## I. INTRODUCTION

Mobile Ad hoc Network (MANET) is a network of a number of mobile routers and associated hosts, organized in a random fashion via wireless links. During recent years MANET has gained enormous amount of attention and has been widely used for not only military purposes but for search-and-rescue operations, intelligent transportation system, data collection, virtual classrooms and ubiquitous computing [13]. Vehicular Ad hoc network is an application area of ad hoc network that is formed by a collection of vehicles that interacts with other vehicles and roadside units to share traffic information, weather information, parking information, internet access etc [1]. Unlike traditional ad hoc networks that often have very low powered devices with small computing capabilities, VANET can be equipped with large batteries, processing capabilities etc [7]. In addition the nodes of VANET moves with higher speed and the number of nodes forming VANET is also very large [2].

Therefore the issues of VANET also differ significantly from traditional issues of MANET. For the research community, the key areas of research in VANET include frequency bands, physical / link layers standards, routing, IP Addressing, security, data dissemination and development of new applications [1],[2],[4].

Different frequency bands have been proposed for VANET (e.g. DSRC, new ARIB STD-T75). IEEE has also proposed IEEE P1609.1, P1609.2, P1609.2 and P1609.4 for VANET. Because of these multiple standards, the adoption rate of VANET is very slow since each country follows its own specifications [1]. Work is also required to address various physical and link layer issues of VANET (e.g. efficient channel estimation and QoS models etc [1]).

Due to the fast changing topology and scale of VANET, routing protocols of MANET (e.g. AODV, DSR, TORA [13]) that were designed considering low mobility and simple movement patterns are not good enough for VANET [3]. Work is being conducted on designing efficient routing protocols that considers metropolitan area mobility models and realistic speed patterns [3],[5]. Another important issue of VANET that is inherited from MANET is maintaining security and trust management among nodes. Since the nodes of VANET are strangers to each other, maintaining security and privacy along with maintaining a reasonable level of cooperation among nodes is a very challenging issue in VANET.

As far as applications are concerned, VANET has been used to solve a large number of safety (reducing accidents, reducing congestion, emergency braking etc. [1],[2]) and non-safety applications (Electronic Toll Free Collection, Travel and Tourism, Multimedia and Gaming applications etc. [1]).

In this paper we have extended the idea of [7], [8] and proposed a novel approach to traffic routing by establishing a grid environment between cars. Grid Computing is also an emerging discipline and is defined as a flexible, secure and coordinated resource sharing among virtual organizations (individuals, institutions, organizations) [11].

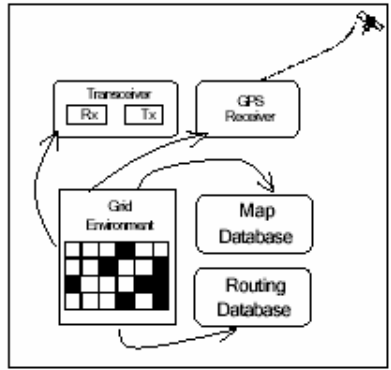

Fig 1: Block diagram of components available at every node

By having a large collection of cars working together in a grid environment, drivers can easily determine the best route (route with minimal congestion) to reach their destination from their current location. Rest of the sections is organized as follows. First we will discuss the related work. Then we will propose our approach to traffic routing. Finally we will conclude the paper with future work.

## II. LITERATURE REVIEW

A huge amount of work is underway to solve traffic problems in metropolitan areas. iTransIT is working on urban traffic management by providing a real time congestion map for city of Dublin[9]. JETS is also working on a prototype implementation to provide a context aware journey time information [11]. This context aware system will index all journey times with the context (time, weather, road usage patterns) in which journey has occurred. FleetNet [6] is an infrastructure less self-organizing traffic information system. Vehicles receive traffic information from other vehicles and analyze the information locally and then transmit information to other vehicles.

VGrid [7],[8] integrates Vehicular Ad hoc network along with Grid Computing technology to solve a large number of traffic related problems like lane merging, ramp metering etc.

In this paper we have extended the idea proposed in [7] and [8] to solve traffic routing problem using grid based vehicular ad hoc network.

## III. PROPOSED APPROACH

We are extending the work of [7] and [8] by having a number of vehicles forming a computational grid to determine the routes for nodes based on any distributed shortest path algorithm [14]. The advantage of using grid computing for VANET is that the approach is scalable as more nodes are added, more computing power is also added. Hence the approach works for a large number of nodes. In addition, if each node will be calculating routes on its own, nodes will be performing redundant calculations. By having a collection of nodes working together to calculate routes to destinations from a single common source, we can avoid a number of redundant calculations.

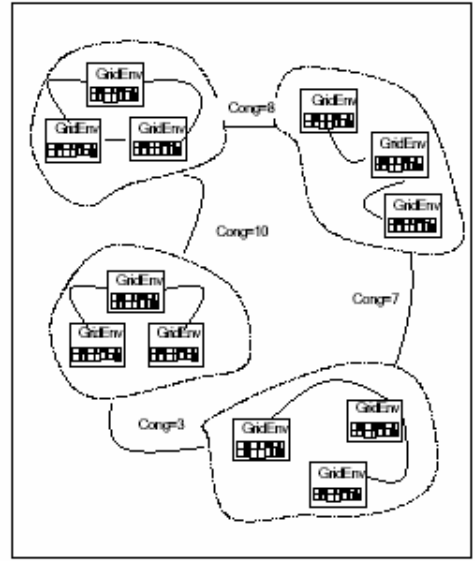

Fig 2: A Grid based approach to Traffic Routing in VANET

In our approach, we are making several simple assumptions. We are assuming that each vehicle is equipped with: a GPS receiver, a digital map, a transceiver, storage and processing capabilities. These simple assumptions are the same as [6]. The GPS receiver will provide the latitude and longitude information that can further be mapped to a particular location using digital map. The digital map also provides a set of routes to reach from a particular source to a destination. Each of the nodes will be running a grid computing environment. Fig 1 describes the block diagram of the components required at every node.

Based on the location information received from GPS, vehicles of near by location will form groups. Members of same group will form a grid and work together to find out feasible routes (minimal congestion) to reach their destination.

## IV. DETAILS OF THE PROPOSED APPROACH

Fig 2 describes the overall approach in detail. Every node will be running a grid-computing environment. Based on the location information received by GPS receiver, vehicles of a particular region will use a clustering algorithm (e.g. k-means) to form a group.

Every group will be assigned a unique group Id. Groups will be interacting with its neighboring groups and sharing information about congestion information in its region. The Congestion information can be obtained by the approach of [7], [8]. As the result of this intergroup communication, a fair indication of congestion in other areas is available to every node. This congestion information will be served as a metric for determining an appropriate path to reach the destination. The nodes of the group will be running a distributed Shortest Path algorithm on their grid environment [14]. This shortest path algorithm will compute shortest paths from a single source to all the destinations.

## V. CONCLUSION

In this paper we have proposed a novel approach based on grid computing for traffic routing. The proposed approach is scalable and based on few simple assumptions, hence easy to be deployed on any environment. The extension to current work includes the design of new routing metrics (e.g. not only congestion but a function of congestion, distance, weather etc.).

ACKNOWLEDGMENT

This research work is supported by 'Center of Research in Ubiquitous Computing', National University of Computer and Emerging Sciences, Karachi, Pakistan